\documentclass[sigplan,nonacm]{acmart}
\usepackage{booktabs} % For formal tables
\usepackage{graphicx}
\usepackage[english]{babel}
\usepackage{hyperref}
\usepackage{dsfont}
\usepackage{amsmath}
\usepackage{latexsym}
\usepackage{graphicx}
\usepackage{listings}
\usepackage{float}
\usepackage{multirow}
\usepackage[scaled]{helvet}
\usepackage[noend]{algorithmic}
\usepackage{mathrsfs}
\usepackage{mathpartir}
\usepackage{mathtools}
\usepackage{dsfont} 
\usepackage{stmaryrd}
\usepackage{url}
\usepackage{textcomp} 
\usepackage[frozencache]{minted}

\usepackage[inline]{enumitem}
\usepackage{pifont}
\usepackage{natbib}
\usepackage{varwidth}
\usepackage{xpatch}
\usepackage{xcolor}

\theoremstyle{definition}

\hypersetup{colorlinks,
  linkcolor=ACMDarkBlue,
  citecolor=ACMPurple,
  urlcolor=ACMDarkBlue,
  filecolor=ACMDarkBlue}
\usepackage{bbding}
\usepackage{tikz}
\usetikzlibrary{shadows, calc, decorations.pathreplacing, tikzmark, positioning,patterns, automata}

\makeatletter
\def\arcr{\@arraycr}
\makeatother

\makeatletter
\newcommand{\mtmathitem}{%
\xpatchcmd{\item}{\@inmatherr\item}{\relax\ifmmode$\fi}{}{\errmessage{Patching of \noexpand\item failed}}
\xapptocmd{\@item}{$}{}{\errmessage{appending to \noexpand\@item failed}}}
\makeatother
\definecolor{shadecolor}{gray}{1.00}
\definecolor{ddarkgray}{gray}{0.75}
\definecolor{darkgray}{gray}{0.30}
\definecolor{light-gray}{gray}{0.87}

\newcommand{\ie}{\emph{i.e.}~}

{\unskip\nobreak\hskip 1em plus 1fil\nobreak$\square$
\parfillskip=0pt%
\endtrivlist}

\graphicspath{{./images/}}
\acmYear{2021}
\acmDOI{} % \acmDOI{10.1145/nnnnnnn.nnnnnnn}
\startPage{1}

\setcopyright{none}

\title{GopCaml: A Structural Editor for OCaml}
\author{Kiran Gopinathan}
\affiliation{%
  \institution{National University of Singapore}
    \country{Singapore}
}

\begin{document}

\begin{abstract}
  This talk presents GopCaml-mode, the first structural editing plugin
  for OCaml.
  We will give a tour of the main plugin features, discussing the
  plugin's internal design and its integration with existing OCaml and
  GNU Emacs toolchains.
\end{abstract}

\maketitle

\section{Introduction}
\label{sec:introduction}
Language-aware editor support can vastly improve the overall user
experience of a programming language.
In this talk, we focus on the task of providing \emph{syntactic}
editor support for OCaml, presenting GopCaml-mode, a plugin for GNU
Emacs, that extends Emacs with support for \emph{syntax-directed}
editing operations.
As it turns out, the particular style of OCaml syntax produces unique
challenges for capturing the structure of programs within the
interfaces of standard editors.

Consider the task of providing structural editing support for GNU
Emacs, one of the most popular editors within the OCaml
community~\cite{ocamlusersurvey2020}.
The base interface of Emacs revolves around a user interacting with a
file by expressing operations in terms of character-based
transformations of text (\ie insert character at cursor).
Plugins then provide additional operations to allow a user to express
\emph{syntax}-based transformations of text in terms of these lexical
operations, \ie delete expression at cursor \emph{delimited by
  braces}, swap statements at cursor \emph{separated by semicolons}.
In the case of OCaml, this approach falls short, as OCaml's syntax
means that the structure of OCaml programs can not consistently be
approximated from a lexical analysis alone --- what denotes the start
and end of a given expression?
% %

While OCaml has a rich tooling ecosystem, existing plugins do not
adequately provide support for structural editing.
Tuareg, Vim-OCaml and VS-code-OCaml-platform, are the main plugins
that provide OCaml-specific language support for the popular editors,
GNU Emacs, Vim and VS-Code respectively. 
While all three plugins provide basic structural editing support,
these functionalities are implemented using basic lexical analysis
alone, and so, by design, are not intended to provide a fully accurate
encoding of OCaml syntax, which results in their inability to support
common OCaml code transformations, such as swapping branches of a
match-statement.
The Merlin~\cite{bour2018merlin} language server extends an editor to
provide additional \emph{semantic} integration with the language, but
does not focus on lower-level \emph{syntactic} editing support.
More recently, the
Rotor~\cite{rowe2019rotor,rowe2018towards,rowe2017rotor} tool is a
framework designed for refactoring OCaml projects, however its focus
is on large-scale whole program transformations rather than local
syntactic transformations as needed for structural editing.

\section{A tour of GopCaml}
\label{sec:tour-gopcaml}
In this section, we highlight a selection of structural editing
operations on OCaml mode that GopCaml-mode supports.

Consider the following snippet of code, where the position of the
user's cursor is denoted by the hollow caret block:

\begin{minted}[escapeinside=@@]{ocaml}
let rec map f xs = match xs with
  @\tikzmark{c1s}@|@\tikzmark{c1e}@ [] -> []
  | x :: xs -> f x :: map f xs
\end{minted}
\begin{tikzpicture}[remember picture,overlay]
  \draw[draw=blue!50, fill=blue!5, fill opacity=0.5, line width=0.5mm]
  ($(pic cs:c1s) - (0.05, 0.1)$) rectangle ($(pic cs:c1e) + (0.05, 0.3)$);
\end{tikzpicture}
\vspace{-1.0em}

\noindent
Now, with this code in hand, what can GopCaml-mode do?

% structural movement
\begin{paragraph}{Structural Navigation}
  One useful operation when editing code is to move the cursor
  relative to the structure of the program, \ie move to the start of
  the match statement \emph{enclosing} the cursor.
  GopCaml-mode explicitly supports such movements --- in this case,
  through a function \texttt{structural-up}, which uses the concrete
  syntax tree (CST) for the current program to reposition the cursor
  accordingly:

\begin{minted}[escapeinside=@@]{ocaml}
let rec map f xs = @\tikzmark{c2s}@m@\tikzmark{c2e}@atch xs with
  | [] -> []
  | x :: xs -> f x :: map f xs
\end{minted}
\begin{tikzpicture}[remember picture,overlay]
  \draw[draw=blue!50, fill=blue!5, fill opacity=0.15, line width=0.5mm]
  ($(pic cs:c2s) - (0.05, 0.1)$) rectangle ($(pic cs:c2e) + (0.05, 0.3)$);
\end{tikzpicture}
\end{paragraph}
\vspace{-1.5em}

% structural transposition
\begin{paragraph}{Structural Transposition}
  Another useful way in which syntactic information can be of use for
  editing is through structural transformations, such as swapping the
  nearest syntactic constructs by the cursor.
  Again, GopCaml-mode supports such operations --- in this case,
  through a function \texttt{structural-transpose}, which uses the
  concrete syntax tree for the current program to modify the program
  text as follows:

\begin{minted}[escapeinside=@@]{ocaml}
let rec map f xs = match xs with
  | x :: xs -> f x :: map f xs
  @\tikzmark{c3s}@|@\tikzmark{c3e}@ [] -> []
\end{minted}
\begin{tikzpicture}[remember picture,overlay]
  \draw[draw=blue!50, fill=blue!5, fill opacity=0.15, line width=0.5mm]
  ($(pic cs:c3s) - (0.05, 0.1)$) rectangle ($(pic cs:c3e) + (0.05, 0.3)$);
\end{tikzpicture}
\end{paragraph}
\vspace{-1.5em}

% structural deletion
\begin{paragraph}{Structural Deletion}
  As a final example, consider the task of deleting entire nodes from
  the syntax tree, \ie delete the branch at the cursor.
  GopCaml-mode supports this transformation through a function
  \texttt{structural-delete}, which again relies on the concrete
  syntax tree for the current program to remove the nearest syntax
  construct to the cursor, leaving the buffer as follows:

\begin{minted}[escapeinside=@@]{ocaml}
let rec map f xs = match xs with
  @\tikzmark{c4s}@|@\tikzmark{c4e}@ x :: xs -> f x :: map f xs
\end{minted}
\begin{tikzpicture}[remember picture,overlay]
  \draw[draw=blue!50, fill=blue!5, fill opacity=0.15, line width=0.5mm]
  ($(pic cs:c4s) - (0.05, 0.1)$) rectangle ($(pic cs:c4e) + (0.05, 0.3)$);
\end{tikzpicture}
\end{paragraph}

\noindent
Other operations supported by GopCaml-mode include:
\begin{itemize}[leftmargin=15pt]
\item \textbf{Structural selection} - select regions using the CST.
\item \textbf{Structural syntax-move} - move nodes around the CST.
\item \textbf{Extract expression} - extract a common sub-expression to
  a let binding.
\item \textbf{Jump to binding/parameter} - move the cursor to the
  nearest let binding/parameter.
\end{itemize}

\section{Under the hood}
\label{sec:under-hood}
% data type
We now present the core logic used by GopCaml-mode to provide
structural editing --- a variant of Huet's zipper~\cite{huet1997zipper},
specialised for navigating syntax trees.
A syntax tree might provide the information needed for structural
editing, but it is fundamentally unsuited for use in an interactive
setting where the tree traversal is iterative, typically being
controlled by the user.
The definition of the zipper used in GopCaml-mode is as follows:
\begin{minted}{ocaml}
type zipper =
  | Top
  | Node of {
      item: t; below: t list; above: t list;
      parent: zipper;
      bounds: text_region;
    }
\end{minted}
An instance of this zipper encodes a path from the root of the program
CST (\ie \mintinline{ocaml}{Top}) to a currently \emph{focused} node
(\mintinline{ocaml}{item}), retaining the structure of the whole tree
by tracking the siblings below and above each node.

The main modification made to Huet's original definition is to extend
each intermediate node with an additional field that captures the
``bounds'' of the current node --- the character range in the original
buffer from the start to the end of the current node.
This change enables a simple interface between the core logic and the
user's editor --- that of simple text ranges, as will be discussed
later in this section.

Finally, the type \mintinline{ocaml}|t| captures a thin wrapper around
OCaml's underlying AST type:
\begin{minted}{ocaml}
type t =
  | Sequence of
      text_region option * t list * t * t list
  | Signature_item of Parsetree.signature_item
  (* ... *)
\end{minted}

\noindent
The full data type has a constructor for each node in the OCaml AST,
along with an additional
\mintinline{ocaml}{Sequence} constructor which provides a generic
encoding of the ``minimal'' information needed to represent an
syntactic node --- this happens to be particularly useful for handling
intermediate operations that produce invalid syntax trees.

As it turns out, using a zipper to track the user's position in the
syntax tree allows for a rather simple and elegant implementation of
the structural operations we saw earlier:
\begin{itemize}[leftmargin=20pt]
\item \textbf{Structural movement} - With this framework, syntax-based
  movement simply boils down to moving the zipper itself and then
  updating the position of the editor's cursor to reflect the change
  in the focus of the zipper.
  For example, in order to move the cursor to the enclosing match
osition  expression as before, the plugin first moves the zipper up the syntax tree by
  replacing it with the value of its parent and then simply updates
  the editor's cursor to move to the start of the newly focused
  element.
\item \textbf{Structural transposition} - Syntax-based transformations
  also neatly fit into this framework.
  For example, in order to transpose the match branches from before,
  the plugin first retrieves the two nodes to be changed as the
  focused element of the zipper and its next sibling (from the
  \mintinline{ocaml}{item} and \mintinline{ocaml}{above} fields
  respectively).
  After swapping them in the zipper, to perform this transformation on
  the program text itself, the plugin passes the text ranges
  corresponding to the swapped elements to the editor, which then
  simultaneously swaps the corresponding characters.
\item \textbf{Structural deletion} - Structural deletion requires more
  care, but still works quite cleanly with zippers.
  In order to delete an element in the zipper, the plugin need only
  remove the currently focused item and replace it with its next
  sibling, and similarly ask the editor to delete the text region
  corresponding to the removed element.
  Finally, to account for the removed characters from the text buffer,
  the zipper must update the text bounds for all subsequent and parent
  nodes accordingly --- by shifting or shrinking them respectively.
\end{itemize}
\vspace{-0.5em}
% structural movement
% structural transposition
% structural deletion

\vspace{-1em}
\section{Editor integration and Future Work}
\label{sec:extending-further}
We have implemented the above zipper-based structural editing
framework as a small OCaml library, using the OCaml compiler
infrastructure to provide the AST definitions and parser.
The GopCaml-mode Emacs plugin builds on this core framework to
integrate it with GNU Emacs, providing functions to persist and cache
the state of the zipper over an editing session and keybindings to run
structural operations using the zipper.
For example, when the user presses a keybinding for a structural
operation, GopCaml-mode then transparently handles the tasks of
building the CST, constructing a zipper at the cursor and performing
the editing operation on the zipper and text buffer.

In the future, we plan to extend support to other editors.
The core editing framework is agnostic to the particular choice of
editor, simply implementing edits in terms of zipper transformations
as above, so it would not be too challenging to integrate this into
editors such as Vim and VS-code.
Finally, looking further out, we hope to extend this framework
further, moving from the \emph{syntactic} to the \emph{semantic} - for
instance, using typing information to guide transformations.

\bibliography{references}
\end{document}